\documentstyle[12pt]{article}

\begin{document}

\renewcommand{\thefootnote}{\fnsymbol{footnote}}


\begin{center}
{\large\bf $A^{(2)}_2$ Parafermions: \\
A New Conformal Field Theory}

\vspace{1cm}

{\normalsize\bf
Xiang-Mao Ding $^{a,b}$ \footnote{E-mail:xmding@maths.uq.edu.au},
Mark D. Gould $^a$ and Yao-Zhong Zhang $^a$ \footnote{E-mail: 
yzz@maths.uq.edu.au}
}          

{\em $^a$ Center of Mathematical Physics, Department of Mathematics, \\
University of Queensland, Brisbane, Qld 4072, Australia.}
\\
{\em $^b$ Institute of Applied Mathematics, \\
Academy of Mathematics and System Sciences; \\
Chinese Academy of Sciences, P.O.Box 2734, 100080, China.} 

\end{center}


\vspace{10pt}

\def\beq{\begin{equation}}
\def\eeq{\end{equation}}
\def\beqa{\begin{eqnarray}}
\def\eeqa{\end{eqnarray}}         
\def\beqas{\begin{eqnarray*}}
\def\eeqas{\end{eqnarray*}}
\def\bea{\begin{array}}
\def\eea{\end{array}}   
\def\bds{\begin{displaymath}}
\def\eds{\end{displaymath}}

\def\nn{\nonumber}
\def\no{\noindent}

\def\bebb{\begin{thebibliography}}
\def\eebb{\end{thebibliography}}      
\def\bbit{\bibitem}

\newcommand {\eqn}[1]{(\ref{#1})}
\newcommand {\eq}[1]{eq.(\ref{#1})}
\newcommand {\eqs}[1]{eqs.(\ref{#1})}
\newcommand {\Eq}[1]{Eq.(\ref{#1})}
\newcommand {\Eqs}[1]{Eqs.(\ref{#1})}
\newcommand {\Label}[1]{\label{#1}}
\newcommand {\eqdef}{\stackrel{\rm def}{=}}

\def\al{\alpha}
\def\bt{\beta}

\def\dl{\delta}
\def\Dl{\Delta}

\def\eps{\epsilon}
\def\vpsn{\varepsilon}

\def\Gm{\Gamma}
\def\gm{\gamma}

\def\lm{\lambda}
\def\Lm{\Lambda}

\def\omg{\omega}
\def\Omg{\Omega}

\def\sgm{\sigma}
\def\Sgm{\Sigma}
\def\vsgm{\varsigma}




\def\tl{\tilde}  
\def\empst{\emptyset}
\def\p{\partial}
\def\nft{\infty}


\def\ot{\otimes}
\def\op{\oplus}

\def\dg{\dagger}
\def\eqv{\equiv}

\def\psd{\psi^{\dg}}
\def\Psd{\psi^{\dg}}
\def\journal#1&#2(#3){\unskip, \sl #1\ \bf #2 \rm(#3) }
\def\andjournal#1&#2(#3){\sl #1~\bf #2 \rm (#3) }
\def\npb#1#2#3{Nucl. Phys. {\bf B#1}, #2 (#3)}
\def\pl#1#2#3{Phys. Lett. {\bf #1B}, (#2) #3}
\def\plb#1#2#3{Phys. Lett. {\bf B#1}, (#2) #3}
\def\prl#1#2#3{Phys. Rev. Lett. {\bf #1},#2 (#3)}
\def\physrev#1#2#3{Phys. Rev. {\bf D#1}, (#2) #3}
\def\prd#1#2#3{Phys. Rev. {\bf D#1} (#2) #3}
\def\ap#1#2#3{Ann. Phys. {\bf #1}, (#2) #3}
\def\prep#1#2#3{Phys. Rep. {\bf #1}, (#2) #3}
\def\rmp#1#2#3{Rev. Mod. Phys. {\bf #1}, (#2) #3}
\def\cmp#1#2#3{Commun. Math. Phys. {\bf #1}, (#2) #3}
\def\cqg#1#2#3{Class. Quant. Grav. {\bf #1}, (#2) #3}
\def\mpl#1#2#3{Mod. Phys. Lett. {\bf A#1}, (#2) #3}
\def\lmp#1#2#3{Lett. Math. Phys. {\bf #1}, (#2) #3}

\begin{abstract}
A new parafermionic algebra associated with the homogeneous 
space $A^{(2)}_2/U(1)$ and its corresponding 
$Z$-algebra have been recently proposed. In this paper, 
we give a free boson representation of the $A^{(2)}_2$ parafermion 
algebra in terms of seven free fields. Free field realizations of the 
parafermionic energy-momentum tensor and screening currents
are also obtained. A new algebraic structure is discovered, which contains 
a $W$-algebra type primary field with spin two. 
\end{abstract}


\setcounter{section}{0}
\setcounter{equation}{0}
\section{Introduction}

The notion of parafermions \cite{ZaFa} was introduced in the 
context of statistical models and conformal field theory \cite{BPZ}. 
Parafermions generalize the Majorana fermions and have found 
important applications in many areas of physics. From statistical 
mechanics point of view, parafermions  are related to the
exclusion statistics introduced by Haldane~\cite{Hald}. 
In particular, the $Z_k$ parafermion models offer various 
extensions of the Ising model which corresponds to the $k=2$ 
case~\cite{FaZa,ZQiu,GeQi,Yang,Neme,DiQi,FrWe}. 

The category for parafermions (nonlocal operators) is the generalized 
vertex operator algebra \cite{DL,LW}. The $Z_k$ parafermion algebra 
was referred to as $Z$-algebra in \cite{DL,LW}, and it was proved that 
the $Z$-algebra is identical with the $A^{(1)}_1$ parafermions.  

The $Z_k$ parafermions proposed in \cite{ZaFa} are basically 
related to the simplest $A_1^{(1)}$ algebra. Various extensions 
have been considered by many researchers. Gepner proposed a
parafermion algebra associated with any given untwisted affine 
Lie algebra ${\cal G}^{(1)}$~\cite{Gepn,Gepn2}, which has been
subsequently used in the study of $D$-branes. The operator product
expansions (OPEs) and the corresponding $Z$-algebra of the 
untwisted parafermions were studied in \cite{DFSW}, and a 
$W_3$-algebra was constructed from the $SU(3)_k$ parafermions. 
In \cite{WaDi} a $W_5$-algebra was constructed by using the 
$SU(2)_k$ parafermions.

More recently, Camino et al extended the $Z_k$ parafermion 
algebra and investigated graded parafermions associated with 
the $osp(1|2)^{(1)}$ Lie superalgebra~\cite{CRS}. The central 
charge of the graded parafermions algebra is $c=-\frac{3}{2k+3}$, 
which implies that for $k$ a positive integer, $c$ is always 
negative. Thus the graded parafermion theory is not unitary. 

In \cite{DGZ2}, we found a new type of nonlocal currents 
(quasi-particles), which were referred to as twisted parafermions. 
The nonlocal currents take values in the homogeneous space 
$A^{(2)}_2/U(1)$. To distinguish our theory from the twisted 
versions of~\cite{ZaFa,DL,LW}, in this paper we refer our nonlocal 
currents to as $A^{(2)}_2$ parafermions. 

The $A^{(2)}_2$ parafermion system contains a boson-like spin-$1$ 
field and six nonlocal fields with fractional spins 
$1-\frac{1}{4k}$ and $1-\frac{1}{k}$, and leads to a new 
conformal field theory which is different from the known ones. 
Let us remark that, while fields with conformal dimensions of 
$1-\frac{1}{4k}$ and $1-\frac{1}{k}$ 
also appeared in the graded parafermion algebra \cite{CRS}, our 
$A^{(2)}_2$ parafermion theory is quite different. 
In particular, our theory is unitary and has a different 
central charge. 

In this paper, we investigate the theory further. We give a free 
field representation of the $A^{(2)}_2$ parafermion theory in 
terms of seven free bosons. We also obtain the free field 
realization of the highest weight state and screening 
currents. We discover a new algebraic structure which contains 
a spin-$2$ primary field. This is similar to a $W$-algebra 
structure but now the spin is two. 

The layout of the paper is as follows. In section $2$, we briefly 
review the $A^{(2)}_2$ parafermion algebra obtained in 
\cite{DGZ2}, which will be extensively used in this paper. 
In section $3$, we give some results concerning the 
$A^{(2)}_2$ parafermion Hilbert space. In 
section $4$, we give a free field representation of the 
$A^{(2)}_2$ parafermion currents, highest weight 
state and screening currents. We present a new 
$W$-algebra structure in section $5$.

\setcounter{section}{1}
\setcounter{equation}{0}
\section{Twisted parafermions: a brief review}

The parafermion current algebra proposed in 
\cite{DGZ2} is related to the twisted affine 
Lie algebra $A^{(2)}_2/U(1)$. This $A^{(2)}_2$ parafermion theory 
is not an extension of the usual $Z_k$ parafermions, but a twisted 
version of the Gepner construction~\cite{Gepn}. To begin with, let 
$g$ be a simple 
finite-dimensional Lie algebra and $\sgm $ be an automorphism of 
$g$ satisfying $\sgm ^r =1$ for a positive integer $r$, then $g$ 
can be decomposed into the form \cite{Kac}: 
\beq
g=\oplus_{j\in {\bf Z}/r{\bf Z}}~~ g_j, 
\eeq

\no where $g_j$ is the eigenspace of $\sgm $ with eigenvalue 
$e^{2j{\pi} i/r}$, and $[g_i, g_j]\subset g_{(i+j)~mod~r}$~, 
then $r$ is called the order of the automorphism. For $A^{(2)} _2$ 
algebra, we have $g=A_2$ so that
\beq
A_2=g_0\op g_1,
\eeq

\no where $g_0$ is the fixed point subalgebra of $A_2$ under the 
automorphism 
and $g_1$ is the five dimensional representation of $g_0$. $g_0$ 
and $g_1$ satisfy $[g_i, g_j]\subset g_{(i+j)~mod ~2}$. 
We associate a generating parafermion to a vector in the root 
lattice $M$ mod $kM_L$, where $M_L$ is the long root lattice and 
$k$ is a constant identified with the level of the corresponding 
twisted affine algebra. Generating parafermions 
are defined by projecting out the affine currents corresponding to the
Cartan subalgebra \cite{Gepn}. Denote by $\psi _{a}$ the $A^{(2)}_2$ 
generating parafermions with $a=\tl{0},~\pm \al,~\pm {\tl{\al}},~\pm 
\tl{\al _2}$ and $\tl{\al _2}=\al + {\tl{\al}}$, where ${\tl{0}}^2 = 
\tl{0}\cdot\al = \tl{0}\cdot\tl{\al} = \tl{0}\cdot\tl{\al_2} = 0$  and
$\al ^2={\tl{\al}} ^2=\al \cdot {\tl{\al}}=1$, ${\tl{\al _2}}^2=4$. 
Then $\psi _{\pm \al}$ are currents in $g_0$ and $\psi_{\tl{0}},~
\psi_{\pm\tl{\al}},~\psi_{\pm \tl{\al _2}}$ are currents in
$g_1$. Note that $\tl{0}$ is a "null" root vector 
in the sense that both its length and its dot product with all
other root vectors are zero. So parafermion currents associated
with $A_2^{(2)}$ live in either Ramond sector (i.e. the currents in $g_0$)
or Neveu-Schwarz sector(i.e. the currents in $g_1$). 

Due to the 
mutually semilocal property between two parafermions, the radial 
ordering products are multivalued functions. So we define the 
radial ordering product of (generating) $A^{(2)}_2$ 
parafermions (TPFs) as
\beqa
\psi _{a}(z)\psi _{b}(w)(z-w)^{a\cdot b/2k}
=\psi _{b}(w)\psi _{a}(z)(w-z)^{a\cdot b/2k}, 
\label{eq:rlrr}
\eeqa
where $a=\tl{0},~\pm \al,~\pm {\tl{\al}},~\pm \tl{\al _2}$. 
{}From the above definition, we have 
\beq
 T_{\psi}(z)\psi _{a}(w)=\frac{\Dl _a}{(z-w)^2}\psi _{a}(w)
+\frac{1}{z-w}\partial \psi _{a}(w)+\ldots.
\label{eq:Tpsi}
\eeq
where $T_{\psi}$ is the energy-momentum tensor of the theory, and 
$\Dl _a=1-\frac{a^2}{4k}$ is the conformal dimension of 
current $\psi _{a}$. It follows that the conformal dimensions of 
the $A^{(2)}_2$ parafermion  currents are 
$1~(a=\tl 0),\;1-\frac{1}{4k}~(a=\pm \al,\pm {\tl \al})$ 
and $1-\frac{1}{k}~(a=\pm {\tl \al _2})$, for a given level $k$. 
The values of the dimensions determine the leading singularity 
in the OPE of the currents. In fact, we  have 
\beqa
\psi _{a}(z)\psi _{b}(w)
&=&(z-w)^{\Dl _{a+b} -\Dl _a -\Dl _b}\left(
\varepsilon _{a,b}\psi _{a+b}(w)+\cdots, \right),\\ 
\psi _{a}(z)\psi _{-a}(w)
&=&(z-w)^{-2\Dl _a}I_0(z)+\cdots,\label{I0}
\eeqa

\no where $\varepsilon _{a,b}$ are the constants, $I_0(z)$ will be
determined below and 
$\Dl _{a}$,~$\Dl _{b}$ and $\Dl _{a +b}$ are the conformal 
dimensions of $\psi _{a}$,~$\psi _{b}$ and 
$\psi _{a+b}$, respectively. If 
$\Dl _{a+b} -\Dl _{a} -\Dl _{b}\geq 0$, 
then two operators $\psi _{a}$,~$\psi _{b}$ 
commute. The nontrivial OPEs are given by 
$\Dl _{a+b} -\Dl _a -\Dl _b < 0$. For example, 
\beq
\Dl _{\tl{\al _2}}-\Dl _{\tl \al}- \Dl _{\al}
=1-\frac{1}{k}-2\left(1-\frac{1}{4k}\right)=-1-\frac{1}{2k}, 
\eeq

\no In the OPE form, this corresponds to 

\beq
\psi _{\al}(z)\psi _{\tl{\al}}(w)
=\frac{\varepsilon _{\al,\tl{\al}}}{(z-w)^{(1+1/2k)}}
\psi _{\tl{\al _2}}(w)+\cdots,
\eeq

\no or equivalently,

\beq
\psi _{\al}(z)\psi _{\tl{\al}}(w)(z-w)^{1/2k}
=\frac{\varepsilon _{\al,\tl{\al}}}{z-w}
\psi _{\tl{\al _2}}(w)+\cdots.
\eeq
Similarly, 

\beq
-\Dl _{\al}-\Dl _{-\al}=-2\left(1-\frac{1}{4k}\right)
=-2+\frac{1}{2k}.
\label{eq:1-1}
\eeq

\no From the dimensional analysis, the conformal dimension of 
$I_0(z)$ on the r.h.s. of the 2nd equation of (\ref{I0}) is zero. It
follows that $I_0(z)$ is a constant, which will be set to 1 in the
sequel.

We have seen that $\psi _{\tl {0}}$ is a spin-$1$ 
current in $g_1$ that corresponds to the ``null" root vector $\tl{0}$. 
Keeping in mind the fact 
that $\psi_{\pm\al}\in g_0$ and $\psi_{\pm\tl{\al}} \in g_1$, we have
$\psi_\al\psi_{-\tl{\al}}\sim\psi_{\al-\tl{\al}}\in g_1$,~
$\psi_{\tl{\al}}\psi_{-\al}\sim\psi_{\tl{\al}-\al}\in g_1$,~
$\psi_{\tl{0}}\psi_{\al}\sim\psi_{\tl{0}+\al}\in g_1$ and
$\psi_{\tl{0}}\psi_{\tl{\al}}\sim\psi_{\tl{0}+\tl{\al}}\in g_0$.
{}From the dimensional analysis and noting that $(\tl{0}+\al)^2=1=(\tl{0}
+\tl{\al})^2$ and $(\al-\tl{\al})^2=0=(\tl{\al}-\al)^2$, we have the
identifications: $\psi_{\tl{0}+\al}\sim\psi_{\tl{\al}}$,~
$\psi_{\tl{0}+\tl{\al}}\sim\psi_\al$,~ 
$\psi_{\al-\tl{\al}}\sim\psi_{\tl{0}}$ and
$\psi_{\tl{\al}-\al}\sim\psi_{\tl{0}}$. Thus we have 
\beqas
\psi _{\al}(z)\psi _{-\tl{\al}}(w)(z-w)^{-1/2k}
&=&\frac{\varepsilon _{\al,-\tl{\al}}}{z-w}
\psi _{\tl{0}}(w)+\cdots, \\
\psi _{\tl {0}}(z)\psi _\al(w)
&=&\frac{\varepsilon _{{\tl {0}},\al}}{z-w}\psi _{\tl{\al}}(w)+\cdots,\\
\psi _{\tl {0}}(z)\psi _{\tl{\al}}(w)
&=&\frac{\varepsilon _{\tl{0},\tl{\al}}}{z-w}\psi _{\al}(w)+\cdots.
\eeqas

\no We see that $\psi _{\tl {0}}(z)$ 
is a spin-$1$ primary field which transforms the fields in the Ramond 
sector to those in the Neveu-Schwarz sector or vice versa. This is easy to 
understand from the theory of the $A_2$ current algebra~\cite{DGZ1}. 

Summarizing, we may write the OPEs in the general form:
\begin{eqnarray}
\psi _{\mu}(z)\psi _{\nu}(w)(z-w)^{\mu\cdot \nu /2k}
&=&\frac{\delta _{\mu+\nu,0}}{(z-w)^2}
+\frac{\varepsilon _{\mu,\nu}}{z-w}
\psi _{\mu +\nu}(w)+\cdots, \nn\\
\psi _{\tl \mu}(z)\psi _{\tl {\nu}}(w)(z-w)^{\tl \mu\cdot \tl {\nu }/2k}
&=&\frac{\delta _{\tl{\mu}+ \tl {\nu},0}}{(z-w)^2}
+\frac{\varepsilon _{\tl{\mu},\tl{\nu}}}{z-w}
\psi _{\tl{\mu} +\tl{\nu}}(w)+\cdots,\nn \\
\psi _{\mu}(z)\psi _{\tl{\nu}}(w)(z-w)^{\mu \cdot \tl{\nu} /2k}
&=&\frac{\varepsilon _{\mu, \tl{\nu}}}{z-w}
\psi _{\mu + \tl{\nu }}(w)+\cdots, \label{parafermion}
\end{eqnarray}
where $\mu,\;\nu=\pm\al,~ \tl{\mu},\;\tl{\nu}=\tl{0},\;\pm\tl{\al},\;
\pm\tl{\al}_2$;
$\varepsilon _{\mu,\nu},~ \varepsilon _{\tl{\mu},\tl{\nu}}$ and 
$\varepsilon _{\mu,\tl{\nu}}$ are structure constants. 
For consistency, $\varepsilon _{a,b}$ must have 
the properties: $\varepsilon _{a,b}=
-\varepsilon _{b,a}=-\varepsilon _{-a,-b}
=\varepsilon _{-a,a+b}$ and $\varepsilon _{a,-a}=0 $.  
The following structure constants have been found in \cite{DGZ2}: 
\beq
\varepsilon _{\tl{\al},-\tl{\al _2}}
=\varepsilon _{\al,\tl{\al}}
=\varepsilon _{-\al,\tl{\al _2}}
=\frac{1}{\sqrt{k}}, \hskip 0.3cm  
\varepsilon _{\al,-\tl{\al}}
=\varepsilon _{\tl{0},\tl{\al}}
=\sqrt{\frac{3}{2k}},
\label{eq:const}
\eeq

\no and other constants can be obtained by using the symmetry 
properties. The above structure constants are compatible with 
the associativity requirement of the OPEs. This is seen as
follows. Let $A_{a}$ and $B_{b}$ be two arbitrary 
functions of the $A^{(2)}_2$ parafermions with charges $a$ and $b$, 
respectively. We write their OPEs as 

\beq
A_{a}(z)B_{b}(w)(z-w)^{a\cdot b/2k}=
\sum_{n=- [h_A+h_B ] }^{\infty} [A_{a}B_{b}]_{-n}(w)(z-w)^n,
\eeq  

\noindent where $ [ h_A ] $ stands for the integral part of the 
dimension of $A$. Hence we have 
$[ A_{a}B_{b}]_n(w)=\oint_w \ dz \ A_{a}(z)B_{b}(w)
(z-w)^{n-1+a\cdot b/2k}$. 
It is easy to find the following relation between the 
three-fold radial ordering products

\beqa
&&\left\{\oint_w \ du \oint_w \ dz \ R(A(u)R(B(z)C(w)))\right.\nn\\
&&-\oint_w \ dz \oint_w \ du \ (-)^{a\cdot b/2k}R(B(z)R(A(u)C(w)))\nonumber\\
&&-\left. \oint_w \ dz \oint_z \ du \ R(R(A(u)B(z))C(w))\right\} \nn\\
&&(z-w)^{p-1+b\cdot c/2k}(u-w)^{q-1+a\cdot c/2k}(u-z)^{r-1+a\cdot b/2k}=0,  
\label{eq:threef}
\eeqa

\noindent where integers $p,\;q,\;r$ are in the regions: $-\infty< \,p\, 
\leq  [h_B+h_C ] ,\;\;-\infty<\,q\,\leq \, [h_C+h_A ] ,\;\;-\infty<\,r\,
\leq\, [h_A+h_B ] $; $a,\;b,\;c$ are parafermionic charges of the 
fields $A,\;B$ and $C$, respectively. 
Performing the binomial expansions, we obtain the following identity: 
 
\beqa
&&\sum_{i=p}^{ [h_B+h_C ] }C_{r-1+a\cdot b/2k}^{(i-p)}
[A [BC ]_i ]_{Q-i}(w)+(-)^r\sum_{j=q}^{ [h_C+h_A ] }
C_{r-1+a\cdot b/2k}^{(j-q)}
[B [AC ]_j ]_{Q-j}(w)\nonumber\\
&&=\sum_{l=r}^{ [h_B+h_A ] }(-)^{(l-r)}C_{q-1+a\cdot c/2k}^{(l-r)}
[ [AB ]_lC ]_{Q-l}(w),
\label{eq:Jacobi}
\eeqa

\no where $ Q=p+q+r-1$. This identity is the twisted 
Jacobi identity of the parafermion algebra.  

Now the structure constants of the currents can be derived from the 
(\ref{eq:Jacobi}). For example, we calculate the OPEs of 
$[\psi _{a}\psi_{-a}]_0$ with $\psi _{a}$ and 
$[\psi _{b}\psi_{-b}]_0$. Setting $Q=p=2,\;q=1$ and $r=0$ in 
(\ref{eq:Jacobi}), and comparing with (\ref{eq:Tpsi}), 
then we have 

\beq
\sum _a \vpsn _{a, b}\vpsn _{-a,a+b}=\frac{6-b^2}{k},\hskip 0.5cm
\sum _a a\cdot b=0, \hskip 0.5cm \sum _a (a\cdot b)^2=12b^2.
\eeq

\no The structure constants (\ref{eq:const}) satisfy all 
these equations. 

A special case of the parafermion OPEs is 
\beqa
\psi _a(z)\psi _{-a}(w)(z-w)^{-a^2 /2k}
&=&\frac{1}{(z-w)^2}+\left(2+\frac{24(1-\Dl _a)}{a^2}\right)t_a+\cdots, 
\nn\\
&=&\frac{1}{(z-w)^2}+\frac{2(k+3)}{k}t_a+\cdots
\eeqa

\no where terms $t_a$ in the right hand side
are fields with conformal dimension $2$. 
{}From the conformal field theory, they should be related to the 
energy-momentum tensor. Indeed, if we define,   
\beq
T_{\psi}(z)=\sum _a t_a
\eeq
then
\beq
  T_{\psi}(z)T_{\psi}(w)=
     \frac{c_{\psi}/2}{(z-w)^4}+\frac{2T_{\psi}(w)}{(z-w)^2}
  +\frac{\partial T_{\psi}(w)}{z-w}+\ldots,
\eeq
 
\no where 
\beq
c_{\psi}=2\sum _a \frac{a^2\Dl _a}{2a^2+24(1-\Dl _a)}
= 7-\frac{24}{k+3}=\frac{8k}{k+3}-1, 
\label{eq:cpsi}
\eeq
is the central charge of the $A^{(2)}_2$ parafermion theory.

\setcounter{section}{2}
\setcounter{equation}{0}
\section{$A^{(2)}_2$ parafermion Hilbert space}

we now analyze the structure of the Hilbert representation 
${\cal H}$ space of the parafermion theory. 
For every field in the parafermion theory there are a pair of charges 
$(\lambda, \bar{\lambda})$, which take values in the weight lattice. 
We denote such a field by $\phi _{\lambda, \bar{\lambda}}(z,\bar{z})$ 
\cite{ZaFa,Gepn,DFSW}. 
Let ${\cal H}_{\lm,\bar{\lm}}$ be the subspace of ${\cal H}$ with 
the indicated charges. Then ${\cal H}$ is the direct sum of the 
form:

\beq
{\cal H}=\op {\cal H}_{\lm,\bar{\lm}}
\eeq  

The non-locality $\gamma$ of two fields $\phi _{\lm, \bar{\lm}}$ 
and $\phi_{\lm', \bar{\lm}'}$ is defined by 

\beq
\gamma \left(\phi _{\lm, \bar{\lm}}, \phi _{\lm', \bar{\lm}'}  \right)
=\frac{1}{2k}\left(\lm\cdot \lm' -\bar{\lm} \cdot \bar{\lm}' \right).
\eeq  

\no Note that the parafermionic current 
$\psi _b(z)$ has the left charge 
$\lambda=b$ and the right charge $\bar{\lambda}=0$. So for the currents 
with zero right charges, the exponent is, 

\beq
\gamma \left(\psi _{a}, \psi _{b}  \right)
=\frac{a\cdot b}{2k}, 
\eeq 

\no which is exactly the exponent appeared in (\ref{parafermion}). 
If we rewrite the (\ref{parafermion}) as

\beq
\psi _a(z)\psi _b(w) (z-w)^{a\cdot b/2k}
\equiv \sum _{n=-2} ^{\infty}(z-w)^n [\psi _a\psi _b]_{-n},
\label{eq:Par}
\eeq
  
\no then we have $
 \left[ \psi _{a}\psi _{b} \right] _l=0\;(l\geq 3)$, $
 \left[ \psi _{a}\psi _{b} \right]_2=\delta _{a+b,0}$ and $
 \left[\psi _{a}\psi _{b} \right]_1={\varepsilon _{a,b}}\psi _{a +b}$. 

The OPE of $\psi _{a}$ with 
$\phi _{\lambda, \bar{\lambda}}(z,\bar{z})$ is given by 
 
\beq
\psi _{a}(z)\phi _{\lambda, \bar{\lambda}}(w,\bar{w})
=\sum _{m=-\infty} ^{\infty}(z-w)^{-m-1-a\cdot\lambda/2k}
A_m ^{a,\lambda}
\phi _{\lambda, \bar{\lambda}}(w,\bar{w}),
\eeq

\noindent where $m \in {\bf Z}$ (Ramond sector) for $a=\pm \al$ 
and $m \in {\bf Z+\frac{1}{2}}$ (Neveu-Schwarz sector) for 
$a={\tl {0}},~\pm{\tl \al},~\pm{\tl \al _2}$. $A_m^{a,\lm}$
are modes of the nonlocal parafermion field $\psi_a(z)$ on
$\phi_{\lm,\bar{\lm}}$, and its action 
on $\phi _{\lambda, \bar{\lambda}}(z)$ 
is defined by the integration

\beq
A_m ^{a,\lambda}\phi _{\lambda, \bar{\lambda}}(w,\bar{w})
=\oint _{c_w}\;dz\;(z-w)^{m+a\cdot\lambda/2k}
\psi _{a}(z)\phi _{\lambda, \bar{\lambda}}(w,\bar{w}),
\eeq

\noindent where $c_w$ is a contour around $w$. From the OPEs,
we know that the dimension of 
$A_m ^{a, \lm}\phi _{\lm, \bar{\lm}}(w,\bar{w})$ is 

\beq
\Dl (A_m ^{a, \lm}\phi _{\lm, \bar{\lm}})=h-m-\frac{a\cdot\lm}{2k}
-\frac{a^2}{4k}, 
\eeq

\no where $h$ is the dimension of the parafermionic field 
$\phi _{\lm, \bar{\lm}}(w,\bar{w})$. We define 
$[\psi _a\psi _b]_{-n;m}^{\lambda}$ by 

\beq
[\psi _a\psi _b]_{-n;m}^{\lambda}\phi _{\lambda, \bar{\lambda}}(w,\bar{w})
=\oint _{c_w}\;dz\;(z-w)^{m+n+(a+b)\cdot\lambda/2k}
[\psi _a\psi _b]_{-n}(z)\phi _{\lambda, \bar{\lambda}}(w,\bar{w}),
\eeq

\no Following the standard procedure in conformal field theory (for 
the parafermion theory, see~\cite{ZaFa}), we multiply the last 
equation on both sides 
by $z^{m+a\cdot\lambda/2k}w^{n+b\cdot\lambda/2k}$ and integrate
the resulting equation by choosing appropriate contours, to
get

\beq
\sum_{l=0} ^{\nft}C_{a\cdot b/2k}^{(l)}[A_{m-l}^{a,b+\lm}A_{n+l}^{b,\lm}
-A_{n-l}^{b,a+\lm}A_{m+l}^{a,\lm}]
=\left( m+\frac{a\cdot\lm}{2k} \right)\dl _{a+b,0}\dl _{m+n,0}
+\varepsilon _{a,b}A_{m+n}^{a+b,\lm},
\label{eq:A-A}
\eeq

\no where we have chosen the powers of $z$ and $w$ such that the 
integrands are single valued, and the binomial coefficients 
$C_x^{(l)}$ are

\beq
C_x^{(l)}=\frac{(-)^lx(x-1)
\cdots (x-l+1)}{l!}
\eeq
 
\no and $C_0^{(0)}=C_n^{(0)}=C_{-1}^{(l)}=1,\;C_p^{(l)}=0$, 
for $l>p>0$. Similarly, we get

\beqas
&&\sum_{l=0} ^{\nft}C_{-1+a\cdot b/2k}^{(l)}[A_{m-l}^{a,b
+\lm}A_{n+l}^{b,\lm}
+A_{n-l}^{b,a+\lm}A_{m+l}^{a,\lm}]  \\
&&~~=\frac{1}{2}\left( m+1+\frac{a\cdot\lm}{2k} \right)
\left( m+\frac{a\cdot\lm}{2k} \right)\dl _{a+b,0}\dl _{m+n,0} \\
&&~~~~~~+\left( m+1+\frac{a\cdot\lm}{2k} \right)
\varepsilon _{a,b}A_{m+n}^{a+b,\lm},
+[\psi _a\psi _b]_{0,m+n}^{\lambda}.
\eeqas

\no A special case of this identity is 

\beqa
&&\sum _a\sum_{l=0} ^{\nft}C_{-1-a^2/2k}^{(l)}
[A_{m-l}^{a,\lm-a}A_{n+l}^{-a,\lm}
+A_{n-l}^{-a,a+\lm}A_{m+l}^{a,\lm}]  \nn\\
&&~~=\sum_a \frac{1}{2}\left( m+1+\frac{a\cdot\lm}{2k} \right)
\left( m+\frac{a\cdot\lm}{2k} \right)\dl _{m+n,0}
+\frac{2(k+3)}{k}L_{m+n}^{\lambda}, 
\label{eq:A-T}
\eeqa
where $L_{m+n}^{\lambda}$ represents 
the action of the energy-momentum tensor
$L_{m+n}$ on the field $\phi _{\lambda, \bar{\lambda}}$. 
Using the above identity, we can show that $L_{m}$ satisfies the 
Virasoro algebra with central charge $c_{\psi}$.
We refer the above identity to as $A^{(2)}_2$ $Z$-algebra. 
Generally, we have  

\beqa
&&\sum _{l=0} ^{\infty}C_{-p-1+a\cdot b/2k} ^{(l)}
 \left[A_{m-l-p+q} ^{a,\lambda+b}A_{n+l+p-q} ^{b,\lambda}
+(-1)^p A_{n-l-q-1} ^{b,\lambda+a}A_{m+l+q+1} ^{a,\lambda}\right]
\nonumber\\
&&=C_{m+q+1+a\cdot b/2k} ^{(p+2)}\delta _{a,-b}\delta _{m,-n}
+ C_{m+q+1+a\cdot b/2k} ^{(p+1)}\varepsilon _{a,b}
A^{a+b,\lambda} _{m+n}\nn\\
&&~~~+\sum _{r=0} ^{\infty}C^{(p-r)} _{m+q+1+a\cdot b/2k}
 [\psi _{a}\psi _{b}]_{-r,m+n} ^{\lambda},~~p=2q~~{\rm or}~~2q+1, 
\eeqa

The above results will be extensively used in the sequel. 
Now let $\phi _{\Lm} ^{\Lm}$ be a state in the 
Hilbert space ${\cal H}$, where $\Lm$ takes value on the weight 
lattices. The condition for $\phi _{\Lm} ^{\Lm}$ to be 
a highest weight state is defined by 
 
\beqa
&& A_n ^{a, \Lm}\phi _{\Lm} ^{\Lm}=0, 
\hskip 0.3cm {\rm for}~~ n>0,~~{\rm or}~~ n=0~~{\rm if}~
   a=\alpha, \tl{\al}, \tl{\al}_2,\\
&& L_0\phi _{\Lm} ^{\Lm}
=\Dl _{\Lm} \phi _{\Lm} ^{\Lm},
\eeqa

\no where $\Dl _{\Lm}$ is the conformal dimension of $\phi _{\Lm} ^{\Lm}$, 
which is found by using the identity (\ref{eq:A-T}) to be 

\beq
\Dl _{\Lm} =\left[\frac{\Lm(\Lm+4)}{4(k+3)}+\frac{\Lm ^2}{12(k+3)}
-\frac{\Lm ^2}{4k}\right]. 
\label{eq:DlLm}
\eeq

\no This result agrees with that obtained from the free field 
calculation given in next section. All other state of ${\cal H}$ 
can be obtained from $\phi _{\Lm} ^{\Lm}$ by applying 
$A_{-n} ^{a, \Lm}$ with 
$n\geq 0$ repeatedly. We define the state 

\beq
\phi _{\lm} ^{\Lm}=\prod _j A_{-n_j}^{a_j,\lm _j} 
\phi _{\Lm} ^{\Lm},~n_j\geq 0.  
\eeq

\no It is easy to show

\beq
\left[L_m, A_{-n} ^{a,\lm} \right]=\left[
(m+1)(\Dl _a -1)+(n-\frac{a\cdot\lm}{2k})\right]A_{m-n} ^{a,\lm},
\eeq

\no from which we find
the conformal dimension of the field $\phi _{\lm} ^{\Lm}$: 

\beq
\Dl _{\lm}^{\Lm}=\Dl _{\Lm}+\frac{\Lm ^2}{4k}-\frac{\lm ^2}{4k}
+ \sum _i n_i, 
\eeq

\no where $\lm= \Lm +\sum _i a_i$, and $n_i\in {\bf Z}$ if $a_i= \mu$, 
or $n_i\in {\bf Z+\frac{1}{2}}$ if $a_i= \tl{\mu}$.

In the usual $Z_k$ parafermion theory, 
highest weight state $\Phi ^l _l$ exists 
only if the condition $l\leq k$ is satisfied. This is also the 
unitarity condition of the representation. To obtain the
unitarity condition for our theory, we define
the hermiticity condition: 

\beq
\left(A^{a,~\Lm}_m\right)^{\dagger} \phi ^{\Lm}_{\Lm+a}
=A^{-a,~a+\Lm} _{-m}\phi ^{\Lm}_{\Lm+a}, 
\hskip 0.5cm k^{\dagger}=k, 
\eeq

If $v _{\Lambda}$ is a vacuum state, i.e 

\beq
A^{a,~\Lm} _nv _{\Lambda}=0, ~~~n>0,
\eeq

\no thus we have the norm 

\beqa
& & (A^{a,\Lm}_{-m}v _{\Lambda},~A^{a,\Lm} _{-m} v _{\Lambda})
= (v _{\Lambda},~A^{-a,a+\Lm}_{m}A^{a,\Lm} _{-m} v _{\Lambda}) \nn\\
& &=(v _{\Lambda},~[A^{-a,a+\Lm}_{m}A^{a,\Lm} _{-m}
-A^{a,-a+\Lm}_{-m}A^{-a,\Lm} _{m}] v _{\Lambda}), 
\nn\\
& &=(v _{\Lambda},\left( m-\frac{a\cdot\Lm}{2k}\right) v _{\Lambda}).
\eeqa
This requires $a\cdot\Lm\leq 2km$ for the norm to be positive.
Considering that the minimal value $m$ can take is $m=1/2$ in the
Neveu-Schwarz sector and $m=1$ in the Ramond sector, we must have
$a\cdot\Lm\leq k$ for the representation to be unitary.

In general, the Verma module with highest weight $\Lm$ and level $k$ 
contains infinite many singular vectors~\cite{Feld,BerFeld},
and thus the representations obtained above are not
irreducible. To determine irreducible modules, an BRST
analysis~\cite{Feld} in the Hilbert space is necessary, which
we will not discuss here. Nevertheless, we will in the next
section give a free field realization of screen currents of the
parafermion theory, which is important ingredient in the BRST
method.
  
\setcounter{section}{3}
\setcounter{equation}{0}
\section{Free field representation of the twisted parafermions}

It was shown in \cite{DGZ2} 
that the twisted affine current algebra $A^{(2)}_2$ allows the 
following representation in terms of the twisted parafermionic 
currents:

\beqa
j^+(z)&=&2{\sqrt k} \psi _{\al} (z) 
e^{\frac{i}{\sqrt{2k}}\phi _0(z)}, 
\hskip 0.6cm
j^-(z)=2{\sqrt k} \psi _{-\al} (z) 
e^{-\frac{i}{\sqrt{2k}}\phi _0(z)}, \nn\\
j^0(z)&=&2{\sqrt {2k}} i\p \phi _0 (z), 
\hskip 1.6cm
J^+(z)=2{\sqrt k} \psi _{\tl{\al}} (z) 
e^{\frac{i}{\sqrt{2k}}\phi _0(z)}, \nn\\
J^-(z)&=&2{\sqrt k} \psi _{-\tl{\al}} (z) 
e^{-\frac{i}{\sqrt{2k}}\phi _0(z)}, 
\hskip 0.6cm
J^{++}(z)=2{\sqrt k} \psi _{\tl{\al _2}}(z) 
e^{i {\sqrt\frac{2}{k}}\phi _0(z)},\nn\\
J^{--}(z)&=&2{\sqrt k} \psi _{-\tl{\al _2}} (z) 
e^{-i {\sqrt\frac{2}{k}}\phi _0(z)},
\hskip 0.6cm 
J^{0}(z)=2{\sqrt {6k}} \psi _{\tl{0}} (z). 
\label{eq:A22}
\eeqa

\no where $\phi _0(z)$ is an $U(1)$ current obeying 
$\phi _0 (z)\phi _0 (w)=-ln (z-w)$, and has the modes 
expansion of $\p \phi _0(z)=\sum _{n\in \bf Z}\phi _n z^{n+1}$. 

On the other hand, we know that the twisted affine current 
algebra $A^{(2)}_2$ allows  a free field representation in 
terms of three $(\bt,~\gm)$ pairs and one $2$-component scalar 
field~\cite{DGZ1}. So to get a free field representation of the 
twisted parafermionic currents, one need to projecting out a $U(1)$ 
current, as is seen from (\ref{eq:A22}), and regarding the 
parafermion currents as operators in the space 
$A^{(2)}_2/U(1)$. This implies that seven independent scalar 
fields are needed to realize the $A^{(2)}_2$ parafermion algebra. 
So, we introduce seven scalar fields, 
$\phi (z)$ and $\xi _j(z)$, $\eta _j (z)~(j=0,~1,~2)$, 
which satisfy the following relations:

\beqas
\xi _i(z)\xi _j(w)=-\dl _{ij}{\it ln}(z-w), \hskip 0.3cm
\eta _i(z)\eta _j(w)=-\dl _{ij}{\it ln}(z-w), \\
\phi (z)\phi (w)=-{\it ln}(z-w).
\eeqas 

\no The modes expansions are 

\beqas
\p \chi (z)=\sum _{n\in \bf Z}\chi _n z^{n+1},~~~\chi=\xi _0,~~~\eta _0
\\
\p \chi (z)=\sum _{n\in {\bf Z+\frac{1}{2}}}\chi _n z^{n+1},~~~
\chi=\xi _j,~~\eta _j~(j=1,~2)~~{\rm or}~~\phi.
\eeqas

\no The conformal dimension of $\psi _{b}$ 
is $1-\frac{b ^2}{4k}$. So we make the ansatz about the free field 
representation of the twisted parafermionic currents:

\beqa
&&\psi _{a}(z)=f_{a}(\xi _i(z),\eta _j(z),\phi (z))
e^{{\sqrt{\frac{a^2}{2k}}}\phi (z)}, \nn\\
&&\psi _{-a}(z)=f_{-a}(\xi _i(z),\eta _j(z),\phi (z))
e^{-{\sqrt{\frac{a^2}{2k}}}\phi (z)},
\eeqa

\def\sq2k{\frac{1}{\sqrt{2k}}}\def\2sq2k{\frac{2}{\sqrt{2k}}}
\def\1sq22k{\frac{1}{2\sqrt{2}k}}

\no where $a=\tl{0},~\al,~\tl{\al},~\tl{\al}_2$, 
the factor $e^{\pm \sqrt{\frac{a}{2k}}\phi (z)}$ will 
contribute $-\frac{a ^2}{4k}$ to the conformal dimension of 
$\psi_{\pm a}(z)$, and $f_{\pm a}(\xi _i(z),\eta _j(z),\phi (z))$ 
are operators with conformal dimension one. 
From dimensional analysis, no term of the form 
$e^{\sqrt{\frac{a^2}{2k}}\phi (z)}$ will 
appear in $f_{\pm a}(\xi _i(z),\eta _j(z),\phi (z))$. Recall that the 
dimensions of $f_{\pm a}(\xi _i(z), \eta _j(z), \phi (z))$ are $1$, 
while the dimensions of $\xi _i(z)$, $\eta _j(z)$, $\phi (z)$ are zero, so 
polynomials of $f_{\pm a}$ are the functions of 
$\p\xi _i(z),~\p\eta _j(z),~\p\phi (z)$. Notice that 
$(\xi _i (z) -i\eta _i (z))(\xi _j (w) -i\eta _j (w))$ have no 
contribution to the OPE. We find, after a 
long and tedious calculation, 

\beqa
f_{\al}(z)&=&\1sq22k \left[ -\al _0 \p \xi _0 (z)
-2\p \xi _1 (z)+\p \xi _2 (z)+\p \phi (z)\right]  
\nn\\
&&~~~~~~~~\times {\rm exp}\{\sq2k \left[-\al _0(\xi _0 (z)-i\eta _0(z))
\right. 
\nn\\
&&~~~~~~~~~~~~~~~~~~~~~~~~~\left.-2(\xi _1 (z)-i\eta _1(z))
+( \xi _2 (z)-i\eta _2(z))\right]\},
\nn\\
f_{-\al}(z)&=&\1sq22k 
\{-\left[(4k+1)\left[\al _0 (\p \xi _0 (z)-i\p\eta _0(z))
+2(\p\xi _1 (z)-i\p \eta _1(z))\right.\right.
\nn\\
&&~~~~~~~~~~~~~~~~~~~~~~~~~\left.-(\p \xi _2 (z)-i\p \eta _2(z))-\p \phi (z)
\right] 
\nn\\
&&~~~~~~~~+i\left. \left[\al _0 \p \eta _0 (z)
+2\p \eta _1 (z)-\p \eta _2 (z)\right]\right] 
\nn\\
&&~~~~~~~~\times {\rm exp}
\{
\sq2k \left[\al _0(\xi _0 (z)-i\eta _0(z))
+2(\xi _1 (z)-i\eta _1(z)) \right.
\nn\\
&&~~~~~~~~~~~~~~~~~~~~~~~~~\left.-( \xi _2 (z)-i\eta _2(z))\right] 
\}
\nn\\
&&~~~~~~~~+3\left[
 \al _0 \p \xi _0 (z)+2\p \xi _1 (z)-\p \xi _2 (z)-\p \phi (z) 
 \right] 
 \nn\\
&&~~~~~~~~\times {\rm exp}\{\sq2k \left[
-(4+\al _0)(\xi _0 (z)-i\eta _0(z))
 \right.
\nn\\
&&~~~~~~~~~~~~~~~~~~~~~~~~~\left.-2(1-\al _0)(\xi _1 (z)-i\eta _1(z))
+3( \xi _2 (z)-i\eta _2(z))
\right] \} 
\nn\\
&&~~~~~~~~+2\left[
 2\p \xi _0 (z)-\al _0\p \xi _1 (z)-\p \xi _2 (z)+\p \phi (z) 
 \right] 
 \nn\\
&&~~~~~~~~\times {\rm exp}\{\sq2k \left[
3(\xi _0 (z)-i\eta _0(z))
-(\al _0 +1)(\xi _1 (z)-i\eta _1(z)) \right.
\nn\\
&&~~~~~~~~~~~~~~~~~~~~~~~~~\left.-(1-\al _0)( \xi _2 (z)-i\eta _2(z))
\right] \} 
\nn\\
&&~~~~~~~~-4\left[
 \p \xi _0 (z)-\p \xi _1 (z)+\al _0\p \xi _2 (z)-2\p \phi (z) 
 \right] 
 \nn\\
&&~~~~~~~~\times {\rm exp}\{\sq2k \left[
-7(\xi _0 (z)-i\eta _0(z))
+(1+3\al _0)(\xi _1 (z)-i\eta _1(z)) \right.
\nn\\
&&~~~~~~~~~~~~~~~~~~~~~~~~~\left.+(3-\al _0 )( \xi _2 (z)-i\eta _2(z))
\right] \} 
\nn\\
&&~~~~~~~~+2{\sqrt 3}i \al _0 \left[
 \p \xi _0 (z)+\p \xi _1 (z)+2\p \xi _2 (z)+\al _0\p \phi (z) 
 \right] 
 \nn\\
&&~~~~~~~~\times {\rm exp}\{\sq2k \left[
-2(\xi _0 (z)-i\eta _0(z))
+\al _0 (\xi _1 (z)-i\eta _1(z)) \right.
\nn\\
&&~~~~~~~~~~~~~~~~~~~~~~~~~\left.+( \xi _2 (z)-i\eta _2(z))
\right] \} \},
\nn\\
f_{\tl {\al _2}}(z)&=&\1sq22k \left[ -\p \xi _0 (z)
+\p \xi _1 (z)-\al _0 \p \xi _2 (z)+2\p \phi (z)\right]  
\nn\\
&&~~~~~~~~\times {\rm exp}\{\sq2k \left[-(\xi _0 (z)-i\eta _0(z))
+(\xi _1 (z)-i\eta _1(z))\right.. 
\nn\\
&&~~~~~~~~~~~~~~~~~~~~~~~~~\left.-\al _0( \xi _2 (z)-i\eta _2(z))\right]\},
\nn\\
f_{\tl \al}(z)&=&\1sq22k\{ \left[ 2 \p \xi _0 (z)
-\al _0 \p \xi _1 (z)-\p \xi _2 (z)+\p \phi (z)\right]
\nn\\
&&~~~~~~~~\times {\rm exp}\{ \sq2k \left[2(\xi _0 (z)-i\eta _0(z))
-\al _0 (\xi _1 (z)-i\eta _1(z))\right. 
\nn\\
&&~~~~~~~~~~~~~~~~~~~~~~~~~\left.-( \xi _2 (z)-i\eta _2(z))\right]\} 
\nn\\
&&~~~~~~~~+2\left[\p \xi _0 (z)
-\p \xi _1 (z)+\al _0 \p \xi _2 (z)-2\p \phi (z)\right] 
\nn\\
&&~~~~~~~~\times {\rm exp}\{\sq2k 
\left[-(1-\al _0)(\xi _0 (z)-i\eta _0(z))
+3(\xi _1 (z)-i\eta _1(z))\right. 
\nn\\
&&~~~~~~~~~~~~~~~~~~~~~~~~~\left.
-(1+\al _0)( \xi _2 (z)-i\eta _2(z))\right]\}\},
\nn\\
f_{\tl 0}(z)&=&\frac{\sqrt 3}{2k}\left\{ 
\left[ -\al _0 \p \xi _0 (z)
-2\p \xi _1 (z)+\p \xi _2 (z)+\p \phi (z)\right] \right. 
\nn\\
&&~~~~~~~~\times {\rm exp}\{\sq2k 
\left[-(2+\al _0)(\xi _0 (z)-i\eta _0(z))
\right. 
\nn\\
&&~~~~~~~~~~~~~~~~~~~~~~~~~\left.-(2-\al _0) (\xi _1 (z)-i\eta _1(z))
+2( \xi _2 (z)-i\eta _2(z))\right]\} 
\nn\\
&&~~~~~~~~+\left[2\p \xi _0 (z)
-\al _0\p \xi _1 (z)- \p \xi _2 (z)+\p \phi (z)\right]  
\nn\\
&&~~~~~~~~\times {\rm exp}\{\sq2k \left[
(2+\al _0)(\xi _0 (z)-i\eta _0(z))\right. 
\nn\\
&&~~~~~~~~~~~~~~~~~~~~~~~~~\left.+(2-\al _0)(\xi _1 (z)-i\eta _1(z))
-2( \xi _2 (z)-i\eta _2(z))\right]\} 
\nn\\
&&~~~~~~~~+\left[  \p \xi _0 (z)
-\p \xi _1 (z)+\al _0 \p \xi _2 (z)-2\p \phi (z)\right] 
\nn\\
&&~~~~~~~~\times 
[ 
{\rm exp}\{   \sq2k \left[
-(1-2\al _0)(\xi _0 (z)-i\eta _0(z))  \right. 
\nn\\
&&~~~~~~~~~~~~~~~~~~~~~~~~~~\left.+5(\xi _1 (z)-i\eta _1(z))
-(2+\al _0)( \xi _2 (z)-i\eta _2(z))
\right] \} 
\nn\\
&&~~~~~~~~~~~~+{\rm exp}\{
\sq2k \left[-5(\xi _0 (z)-i\eta _0(z))
+(1+2\al _0)(\xi _1 (z)-i\eta _1(z))\right. 
\nn\\
&&~~~~~~~~~~~~~~~~~~~~~~~~~~~~~\left.
+(2-\al _0)( \xi _2 (z)-i\eta _2(z))\right]\}
] 
\nn\\
&&~~~~~~~~-\frac{i\al _0}{\sqrt 3}\left. \left( \p \xi _0 (z)
+ \p \xi _1 (z)+2\p \xi _2 (z)+\al _0\p \phi (z)\right) \right\}. 
\nn\\
f_{-{\tl \al}}(z)&=&\1sq22k 
\left\{-\left[(4k+5)
\left[2\al _0 (\p \xi _0 (z)-i\p \eta _0(z))
-\al _0(\p \xi _1 (z)-i\p \eta _1(z))
\right.\right.\right.
\nn\\
&&~~~~~~~~~~~~~~~~~~~~~~~~~\left.
-(\p \xi _2 (z)-i\p \eta _0(z))+\p \phi (z)\right]
\nn\\
&&~~~~~~~~\left. -(2-4\al _0)i \p \eta _0 (z)
+(12-\al_0)i\p \eta _1 (z)-(5+4\al _0)i\p \eta _2 (z)\right] 
\nn\\
&&~~~~~~~~\times {\rm exp}
\{
\sq2k \left[-2(\xi _0 (z)-i\eta _0(z))
+\al _0(\xi _1 (z)-i\eta _1(z)) \right.
\nn\\
&&~~~~~~~~~~~~~~~~~~~~~~~~~
\left.+( \xi _2 (z)-i\eta _2(z))\right] 
\}
\nn\\
&&~~~~~~~~+2\left[
 \al _0 \p \xi _0 (z)+2\p \xi _1 (z)-\p \xi _2 (z)-\p \phi (z) \right] 
\nn\\
&&~~~~~~~~\times {\rm exp}\{
\sq2k \left[
(1-\al _0)(\xi _0 (z)-i\eta _0(z))
-3(\xi _1 (z)-i\eta _1(z)) \right.
\nn\\
&&~~~~~~~~~~~~~~~~~~~~~~~~~\left.+(1+\al _0)( \xi _2 (z)-i\eta _2(z))
\right] 
\} 
\nn\\
&&~~~~~~~~+3\left[
 -2\p \xi _0 (z)+\al _0\p \xi _1 (z)+\p \xi _2 (z)-\p \phi (z) 
 \right] 
 \nn\\
&&~~~~~~~~\times {\rm exp}
\{
\sq2k \left[
2(1+\al _0)(\xi _0 (z)-i\eta _0(z))
\right.
\nn\\
&&~~~~~~~~~~~~~~~~~~~~~~~~~\left.+(4-\al _0)(\xi _1 (z)-i\eta _1(z))
-3( \xi _2 (z)-i\eta _2(z))
\right] 
\} 
\nn\\
&&~~~~~~~~-2\left[
 \p \xi _0 (z)-\p \xi _1 (z)+\al _0\p \xi _2 (z)-2\p \phi (z) 
 \right] 
 \nn\\
&&~~~~~~~~\times 
[
{\rm exp}\
\{
\sq2k \left[
-(1-3\al _0)(\xi _0 (z)-i\eta _0(z))
 \right.
\nn\\
&&~~~~~~~~~~~~~~~~~~~~~~~~~~\left.+7(\xi _1 (z)-i\eta _1(z))
-(3+\al _0 )( \xi _2 (z)-i\eta _2(z))
\right] 
\} 
\nn\\
&&~~~~~~~~~~~~+3~{\rm exp}
\{
\sq2k \left[
-(5-\al _0)(\xi _0 (z)-i\eta _0(z)) \right.
\nn\\
&&~~~~~~~~~~~~~~~~~~~~~~~~~~~~~~~
+(3+2\al _0)(\xi _1 (z)-i\eta _1(z))
\nn\\
&&~~~~~~~~~~~~~~~~~~~~~~~~~~~~~~\left.
+(1-\al _0 )( \xi _2 (z)-i\eta _2(z))
\right] 
\}
] 
\nn\\
&&~~~~~~~~+2{\sqrt 3}i \al _0 \left[
 \p \xi _0 (z)+\p \xi _1 (z)+2\p \xi _2 (z)+\al _0\p \phi (z) 
 \right] 
 \nn\\
&&~~~~~~~~\times {\rm exp}
\{
\sq2k \left[
\al _0 (\xi _0 (z)-i\eta _0(z))
+2(\xi _1 (z)-i\eta _1(z)) \right.
\nn\\
&&~~~~~~~~~~~~~~~~~~~~~~~~~\left.\left.-( \xi _2 (z)-i\eta _2(z))
\right] \} 
\right\},
\nn\\
f_{-{\tl {\al _2}}}(z)&=&\1sq22k 
\left\{
-4\left[(k+1)
\left[(\p \xi _0 (z)-i\p \eta _0(z))
-(\p \xi _1 (z)-i\p \eta _1(z))
\right.\right.\right.
\nn\\
&&~~~~~~~~~~~~~~~~~~~~~~~~~\left.
+\al _0(\p \xi _2 (z)-i\p \eta _2(z))-2\p \phi (z)
\right]
\nn\\
&&~~~~~~~~+i\left.\left[\p \eta _0 (z)
-\p \eta _1 (z)+\al _0 \p \eta _2 (z)\right]\right] 
\nn\\
&&~~~~~~~~\times {\rm exp}
\{
\sq2k \left[(\xi _0 (z)-i\eta _0(z))
-(\xi _1 (z)-i\eta _1(z)) \right.
\nn\\
&&~~~~~~~~~~~~~~~~~~~~~~~~~\left. 
+\al _0( \xi _2 (z)-i\eta _2(z))\right] 
\}
\nn\\
&&~~~~~~~~+2\left[
 -\al _0 \p \xi _0 (z)-2\p \xi _1 (z)+\p \xi _2 (z)+\p \phi (z) \right] 
\nn\\
&&~~~~~~~~\times {\rm exp}
\{
\sq2k \left[
-(6+\al _0)(\xi _0 (z)-i\eta _0(z))
 \right.
\nn\\
&&~~~~~~~~~~~~~~~~~~~~~~~~~\left.
-(2-3\al _0)(\xi _1 (z)-i\eta _1(z))+4(\xi _2 (z)-i\eta _2(z))
\right] 
\}
\nn\\
&&~~~~~~~~-2\left[
 (2-\al _0)i\p \eta _0 (z)-(6-\al _0)i\p \eta _1 (z)
 +(2+4\al _0)i\p \eta _2 (z)
 \right.
 \nn\\
&&~~~~~~~~~~~~~~+(4k+2)\left[
-2(\p\xi _0 (z)-i\p\eta _0(z))
+\al _0 (\p\xi _1 (z)-i\p\eta _1(z)) \right.
\nn\\
&&~~~~~~~~~~~~~~~~~~~~~~~~~~~~~~+\left. (\p\xi _2 (z)-i\p\eta _2(z))\right]
-4(k+2)\p \phi (z) 
\nn\\
&&~~~~~~~~~~~~~~-3(2-\al _0)(\p\xi _0 (z)-i\p\eta _0(z)) 
\nn\\
&&~~~~~~~~~~~~~~\left.+3(2+\al _0) \al _0 (\p\xi _1 (z)-i\p\eta _1(z))
\right]
\nn\\ 
&&~~~~~~~~\times {\rm exp}
\{
\sq2k \left[
-(2-\al _0)(\xi _0 (z)-i\eta _0(z)) \right.
\nn\\
&&~~~~~~~~~~~~~~~~~~~~~~~~~\left.
+(2+\al _0)(\xi _1 (z)-i\eta _1(z)) \right]
\}
\nn\\
&&~~~~~~~~+2\left[
 -2\p \xi _0 (z)+\al _0\p \xi _1 (z)+\p \xi _2 (z)-\p \phi (z) 
 \right]
 \nn\\
&&~~~~~~~~\times {\rm exp}
\{
\sq2k \left[
(2+3\al _0)(\xi _0 (z)-i\eta _0(z))
 \right.
\nn\\
&&~~~~~~~~~~~~~~~~~~~~~~~~~\left.
+(6-\al _0)(\xi _1 (z)-i\eta _1(z))
-4( \xi _2 (z)-i\eta _2(z))
\right] 
\} 
\nn\\
&&~~~~~~~~-\left[
 \p \xi _0 (z)-\p \xi _1 (z)+\al _0\p \xi _2 (z)-2\p \phi (z) 
 \right] 
 \nn\\
&&~~~~~~~~\times
[
 {\rm exp}
\{
\sq2k \left[
-(1-4\al _0)(\xi _0 (z)-i\eta _0(z))
 \right.
\nn\\
&&~~~~~~~~~~~~~~~~~~~~~~~~~~\left.
+9(\xi _1 (z)-i\eta _1(z))
-(4+\al _0 )( \xi _2 (z)-i\eta _2(z))
\right] 
\} 
\nn\\
&&~~~~~~~~~~~~-3~{\rm exp}
\{
\sq2k \left[
-9(\xi _0 (z)-i\eta _0(z))
+(1+4\al _0)(\xi _1 (z)-i\eta _1(z)) \right.
\nn\\
&&~~~~~~~~~~~~~~~~~~~~~~~~~~~~~~~~\left.
+(4-\al _0 )( \xi _2 (z)-i\eta _2(z))
\right]
\} 
\nn\\
&&~~~~~~~~~~~~+ 6~{\rm exp}
\left.
\{
\sq2k 
\left[
-(5-2\al _0)(\xi _0 (z)-i\eta _0(z))
 \right.\right.
\nn\\
&&~~~~~~~~~~~~~~~~~~~~~~~~~~~~~~~~\left.
+(5+2\al _0)(\xi _1 (z)-i\eta _1(z))
-\al _0 ( \xi _2 (z)-i\eta _2(z))
\right] 
\} 
]
\nn\\
&&~~~~~~~~+2{\sqrt 3}i \al _0 \left[
 \p \xi _0 (z)+\p \xi _1 (z)+2\p \xi _2 (z)+\al _0\p \phi (z) \right] 
\nn\\
&&~~~~~~~~\times 
[ 
{\rm exp}
\{
\2sq2k \left[
\al _0 (\xi _0 (z)-i\eta _0(z))
+2(\xi _1 (z)-i\eta _1(z))\right.
\nn\\
&&~~~~~~~~~~~~~~~~~~~~~~~~~\left. -( \xi _2 (z)-i\eta _2(z)
\right]
\}
\nn\\
&&~~~~~~~~~~~~-{\rm exp}
\{
\2sq2k \left[
-2 (\xi _0 (z)-i\eta _0(z))
+\al _0 (\xi _1 (z)-i\eta _1(z))\right. 
\nn\\
&&~~~~~~~~~~~~~~~~~~~~~~~~~~~~\left.\left. 
-( \xi _2 (z)-i\eta _2(z))
\right] 
\} 
]
\right\},
\eeqa

\no where $\al _0 =\sqrt{-2(k+3)}$. It can be checked that
$\psi_a(z)$ indeed satisfy all of the OPE relations of our
parafermion algebra. By the above 
representation of the $A^{(2)}_2$ parafermionic currents, 
we can bosonize the parafermion energy-momentum tensor $T_{\psi}$. 
The result is 

\beqa
T_{\psi}(z)&=&-\frac{1}{2} :\{
(\p \xi _0 (z))^2 + (\p \xi _1 (z))^2 + (\p \xi _2 (z))^2 \nn\\
&&~~~~~~~~+(\p \eta _0(z))^2 +(\p \eta _1(z))^2 +(\p \eta _2(z))^2
+(\p \phi (z))^2 \}: \nn\\
&&-\frac{1}{2\sqrt{2k}} \{
-(1-\al _0)(\p ^2 \xi _0 (z)-i\p ^2 \eta _0(z))
+(1+\al _0)(\p ^2 \xi _1 (z)-i\p ^2 \eta _1(z)) \nn\\
&&~~~~~~~~~~~+\al _0(\p ^2 \xi _2 (z)-i\p ^2 \eta _2(z))
-4\p ^2 \phi (z) \} \nn\\
&&+\frac{1}{\sqrt{k(k+3)}} \left(
\p ^2 \eta _0(z) + \p ^2 \eta _1(z) + \p ^2 \eta _2(z) \right).
\eeqa

\no It is easy to check that the central charge of this 
operator is indeed $c_{\psi}$ given in (\ref{eq:cpsi}).

Using the free 
field realization of the parafermionic currents, we obtain a free 
field representation of the highest weight state $V ^{j} _{j}(z)$ of
the parafermion algebra:

\beqa
V ^{j} _{j}(z)&=&{\rm exp} 
\{
\frac{-j}{2\sqrt{k(k+3)}} \left[
\eta _0(z) +\eta _1(z)+2\eta _2(z)\right] 
\}
\nn\\
&&\times 
{\rm exp} 
\{
\frac{-j}{2\sqrt{3k(k+3)}} \left[
\xi _0(z) +\xi _1(z)+2\xi _2(z)\right]-\frac{ij}{\sqrt{6k}} \phi (z)
\},
\eeqa

\no where $j$ is the spin of the highest weight state. 
 
Now we come to the free field realization of screen currents of
the parafermion algebra. After a long computation, we find the
following two screen operators:

\beqa
S_{+}(z)&=&\sq2k 
\{2\left[- \p \xi _0 (z)+\p\xi _1 (z)-\al _0\p \xi _2 (z)+2\p \phi (z)
\right] 
\nn\\
&&~~~~~~~~\times {\rm exp}
\{
\sq2k \left[-3(\xi _0 (z)-i\eta _0(z))
+(1+\al _0)(\xi _1 (z)-i\eta _1(z)) \right.
\nn\\
&&~~~~~~~~~~~~~~~~~~~~~~~~~\left.+(1-\al _0)(\xi _2 (z)-i\eta _2(z))\right] 
\}
\nn\\
&&~~~~~~~~+\left[
 \al _0 \p \xi _0 (z)+2\p \xi _1 (z)-\p \xi _2 (z)-\p \phi (z) 
 \right] 
 \nn\\
&&~~~~~~~~\times {\rm exp}\{\sq2k \left[
-\al _0(\xi _0 (z)-i\eta _0(z))
 \right.
\nn\\
&&~~~~~~~~~~~~~~~~~~~~~~~~~\left.-2(\xi _1 (z)-i\eta _1(z))
+( \xi _2 (z)-i\eta _2(z))
\right] \} 
\nn\\
&&~~~~~~~~+\left[
 2\p \xi _0 (z)-\al _0\p \xi _1 (z)-\p \xi _2 (z)+\p \phi (z) 
 \right] 
 \nn\\
&&~~~~~~~~\times {\rm exp}\{\sq2k \left[
2(\xi _0 (z)-i\eta _0(z))
-\al _0(\xi _1 (z)-i\eta _1(z)) \right.
\nn\\
&&~~~~~~~~~~~~~~~~~~~~~~~~~\left.-( \xi _2 (z)-i\eta _2(z))
\right] \} \}
\nn\\
&&~~~~~~~~\times {\rm exp}\{\frac{i}{\sqrt{2k}\al _0} \left[
({\sqrt 3}\xi _0 (z)+\eta _0(z))
+({\sqrt 3}\xi _1 (z)+\eta _1(z)) \right.
\nn\\
&&~~~~~~~~~~~~~~~~~~~~~\left.+({\sqrt 3}\xi _2 (z)+\eta _2(z))
\right] \} 
{\rm exp}\{\sq2k \left({\sqrt 3}\phi (z) +i\phi(z) \right)\}
\eeqa

\no and 

\beqa
S_{-}(z)&=&\sq2k 
\{2\left[- \p \xi _0 (z)+\p\xi _1 (z)-\al _0\p \xi _2 (z)+2\p \phi (z)
\right] 
\nn\\
&&~~~~~~~~\times {\rm exp}
\{
\sq2k \left[-3(\xi _0 (z)-i\eta _0(z))
+(1+\al _0)(\xi _1 (z)-i\eta _1(z)) \right.
\nn\\
&&~~~~~~~~~~~~~~~~~~~~~~~~~\left.+(1-\al _0)(\xi _2 (z)-i\eta _2(z))\right] 
\}
\nn\\
&&~~~~~~~~+\left[
 \al _0 \p \xi _0 (z)+2\p \xi _1 (z)-\p \xi _2 (z)-\p \phi (z) 
 \right] 
 \nn\\
&&~~~~~~~~\times {\rm exp}\{\sq2k \left[
-\al _0(\xi _0 (z)-i\eta _0(z))
 \right.
\nn\\
&&~~~~~~~~~~~~~~~~~~~~~~~~~\left.-2(\xi _1 (z)-i\eta _1(z))
+( \xi _2 (z)-i\eta _2(z))
\right] \} 
\nn\\
&&~~~~~~~~-\left[
 2\p \xi _0 (z)-\al _0\p \xi _1 (z)-\p \xi _2 (z)+\p \phi (z) 
 \right] 
 \nn\\
&&~~~~~~~~\times {\rm exp}\{\sq2k \left[
2(\xi _0 (z)-i\eta _0(z))
-\al _0(\xi _1 (z)-i\eta _1(z)) \right.
\nn\\
&&~~~~~~~~~~~~~~~~~~~~~~~~~\left.-( \xi _2 (z)-i\eta _2(z))
\right] \} \}
\nn\\
&&~~~~~~~~\times {\rm exp}\{\frac{-i}{\sqrt{2k}\al _0} \left[
({\sqrt 3}\xi _0 (z)-\eta _0(z))
+({\sqrt 3}\xi _1 (z)-\eta _1(z)) \right.
\nn\\
&&~~~~~~~~~~~~~~~~~~~~~~\left.+({\sqrt 3}\xi _2 (z)-\eta _2(z))
\right] \} 
{\rm exp}\{\sq2k \left({\sqrt 3}\phi (z) -i\phi(z) \right)\}.
\eeqa

The OPEs of these two screen currents with the energy-momentum tensor
and parafermion currents are given by 

\beqa
T_{\psi}(z)S_{\pm} (w)=\p _w \left(\frac{1}{z-w}S_{\pm}(w)\right)
+\ldots,\nn\\
\psi _\al(z)S_{\pm}(w)=\ldots, ~~~~~~
\psi _{\tl \al}(z)S_{\pm}(w)=\ldots, \nn\\
\psi _{-\al}(z)S_{\pm}(w)=\p _w \left(\frac{2\al ^2 _0}
{z-w}{\tl S}_{\pm}(w)\right)+\ldots,\nn\\
\psi _{\tl{\al}_2}(z)S_{\pm}(w)=\ldots,~~~~~~
\psi _{\tl 0}(z)S_{\pm} (w)=\ldots, \nn\\
\psi _{-\tl \al}(z)S_{\pm} (w)=\p _w \left(\mp \frac{2\al ^2 _0}
{z-w}{\tl S}_{\pm}(w)\right)+\ldots, \nn\\
\psi _{-\tl{\al}_2}(z)S_{\pm}(w)=\p _w \left(\mp 
\frac{4\al ^2 _0}{z-w}
{\tl S}_{\pm 2}(w)\right)+\ldots, 
\eeqa

\no where, 

\beqa
{\tl S}_{\pm}(z)=&&{\rm exp}\{\frac{\pm i}{\sqrt{2k}\al _0} 
\left[
({\sqrt 3}\xi _0 (z)\pm\eta _0(z))
+({\sqrt 3}\xi _1 (z)\pm\eta _1(z)) \right.
\nn\\
&&~~~~~~~~~~~~~~~\left.+({\sqrt 3}\xi _2 (z)\pm\eta _2(z))\right] \} 
{\rm exp}\{\sq2k \left({\sqrt 3}\phi (z) \pm i\phi(z) \right)\}, \nn\\
{\tl S}_{2}(z)&=&\frac{1}{\sqrt{2k}}\left[-i(2+\al _0)\xi _0 (z)
-i(2-\al _0)\xi _1 (z) +2i\xi _2 (z)\right] {\tl S}_{+}(z), \nn\\
{\tl S}_{-2}(z)&=&\frac{1}{\sqrt{2k}}\left[-i(2-\al _0)\xi _0 (z)
-i(2+\al _0)\xi _1 (z) +2i\phi (z)\right] {\tl S}_{-}(z), 
\eeqa

\setcounter{section}{4}
\setcounter{equation}{0}
\section{Spin-$2$ Primary field and novel algebraic structure}

In conformal field theory, primary fields are fundamental objects. 
The descendant fields can be obtained from primary fields.

It is well know that, the energy-momentum stress is not primary field
(unless the central charge is zero). The lowest spin of the primary 
field obtained by the Hamiltonian reduction approach is 
three~\cite{FaLy,FORT}. This agrees with the other methods, such as 
high order Casimir, coset model~\cite{BBSS}, free field 
realization~\cite{Frnk}, or the parafermion construction~\cite{DFSW}. 
For more references about $W$-algebras see the 
reviews~\cite{FORT,BoSc}, and their applications in string and 
gravity theory see~\cite{Hull,West}. In the following, we use the 
$A^{(2)}_2$ parafermionic currents to construct a primary field of 
spin two. Define the spin-$2$ currents:

\beq
{\tl w}_2(z)
=\frac{k(4k-1)}{4(2k+3)(k-1)}
\sum _{a=\pm \al, \pm{\tl \al}} [\psi _{a}\psi_{-\tl a}]_0. 
\label{eq:w2}
\eeq

\no The action of ${\tl w}_2(z)$ on $\psi _b(z)$ is given by the 
following OPE:

\beqa
{\tl w}_2(z) \psi _{b}(w)&=&\frac{\Dl _{\tl b}}{(z-w)^2}\psi _{\tl b}(w)
+\frac{1}{z-w}\partial \psi _{\tl b}(w)+\ldots, ~~b=\pm \al, \pm {\tl \al}, 
\nn\\
{\tl w}_2(z) \psi _{\pm \tl {\al _2}}(w)&=& \cdots, \hskip 0.3cm 
{\tl w}_2(z) \psi _{\tl {0}}(w)= \cdots,
\label{eq:w2ps}
\eeqa

\no It should be be understood that ${\tl {\tl b}}=b$ in the first 
line of (\ref{eq:w2ps}). This OPE fixes the normalization 
of ${\tl w}_2(z)$. (\ref{eq:w2ps}) suggests that ${\tl w}_2(z)$ 
behaves similar to a energy-momentum tensor except that it 
transforms $\psi _{b}$ into $\psi _{\tl b}$ and vice versa. 
Recall that $\psi _{b}$ and $\psi _{\tl b}$ have the same 
conformal dimensions. The OPEs of 
${\tl w}_2(z)$ with itself and the stress tensor are

\beqa
{\tl w}_2(z){\tl w}_2(w)&=&\frac{{\tl c}/2}{(z-w)^4}
+\frac{2U(w)}{(z-w)^2}+\frac{\p U(w)}{(z-w)}+\cdots.\nn\\
T_{\psi}(z){\tl w}_2(w)&=&\frac{2{\tl w}_2(w)}{(z-w)^2}
+\frac{\partial {\tl w}_2(w)}{z-w}+\ldots,
\label{eq:w2w2t}
\eeqa

\no where, 

\beq
{\tl c}=\frac{(4k-1)^2}{2(k-1)(2k+3)}
=\frac{(13c_{\psi} +5)^2}{24(c_{\psi} +9)(c_{\psi} -1)}, 
\eeq

\no is the central charge of ${\tl w}_2(z)$, and $c_{\psi}$ is the 
central charge of the $A^{(2)}_2$ parafermionic energy-momentum 
tensor (\ref{eq:cpsi}), and $U(z)$ is a spin two field given by 

\beqa
U(z)&=&\frac{(k+1)(k+3)(4k-1)^2}{4(2k+3)^2(k-1)^2}
\left(
T_{\psi}(z)-\frac{1}{2}\Omg _0(z) 
-\frac{k(k+2)}{(k+1)(k+3)}\Omg _2(z)
\right), \nn\\
\Omg _0(z)&=&\frac{1}{2}\left[\psi _{\tl 0}\psi _{\tl 0}\right]_0 (z), 
\hskip 0.5cm 
\Omg _2(z)=\frac{k}{k+2} 
\left[\psi _{\tl \al _2}\psi _{-{\tl \al _2}}\right]_0 (z). 
\eeqa

\no The OPEs of  $\Omg _0(z)$ 
and $\Omg _2(z)$ with the parafermion currents are 

\beqas
\Omg _0(z) \psi _{\tl {0}}(w)&=&\frac{1}{(z-w)^2}
\psi _{\tl {0}}(w)
+\frac{1}{z-w}\partial \psi _{\tl {0}}(w)+\ldots, \\
\Omg _0(z) \psi _{b}(w)&=&\frac{3}{4k-1}\left(
\frac{\Dl _b}{(z-w)^2}\psi _{b}(w)
+\frac{1}{z-w}\partial \psi _{b}(w)+\ldots  \right),\\ 
\Omg _0(z) \psi _{\pm \tl {\al _2}}(w)&=&..., \\
\Omg _2(z) \psi _{\tl {0}}(w)&=&...,\\
\Omg _2(z) \psi _{b}(w)&=&\frac{k-1}{2k(k+2)}\left(
\frac{\Dl _b}{(z-w)^2}\psi _{b}(w)
+\frac{1}{z-w}\partial \psi _{b}(w)+\ldots \right),\\ 
\Omg _2(z) \psi _{\pm {\tl {\al _2}}}(w)
&=&\frac{\Dl _{\pm \tl \al _2}}{(z-w)^2}\psi _{\pm \tl {\al _2}}(w)
+\frac{1}{z-w}\partial \psi _{\pm \tl {\al _2}}(w)+\ldots,
\eeqas

{}From (\ref{eq:w2w2t}), we see that the field ${\tl w}_2$ introduced 
above is a primary field with conformal spin two. 
The modes expansion of ${\tl w}_2(z)$ is 
\beq
{\tl w}_2(z)=\sum _{n\in {\bf Z+\frac{1}{2}}}{\tl w}_{2,n}z^{-n-2}.
\eeq    
So like the spin-$1$ primary 
field $\psi _{\tl 0}(z)$, ${\tl w}_2(z)$ lives in the Neveu-Schwarz 
sector, and it transforms fields in the Ramond sector into those in 
the Neveu-Schwarz sector or vice versa. If we regard the spin-2 
currents as affine connections, then the energy-momentum tensor 
is a project connection that transforms fields in one sector, while 
field ${\tl w}_2(z)$ is a project connection which 
transforms a field in one sector into a field in a different 
sector. The mode expansions of $\Omg _0(z)$ and $\Omg _2(z)$ are 
\beq
\Omg _0(z)=\sum _{n\in {\bf Z}}\Omg_{0,~n} z^{-n-2}, ~~~~~
\Omg _2(z)=\sum _{n\in {\bf Z}}\Omg_{2,~n} z^{-n-2}, 
\eeq 
\no respectively. So $\Omg_0(z)$ and $\Omg_2(z)$ live in the
Ramond sector.

{}From the above, we see that ${\tl w}_2(z)$, $T(z)$ do not close to 
an algebra. It can be checked that the field $U(z)$ with itself also 
cannot form a closed algebra. But if we decompose the field $U(z)$ 
as above, then the algebra generated by ${\tl w}_2(z)$, $T(z)$, 
$\Omg _0(z)$, $\Omg _2(z)$ is closed, as can be seen from the 
following additional OPEs, 

\beqa
\Omg _0(z)\Omg _0(w)&=&\frac{c_0/2}{(z-w)^4}+\frac{2\Omg _0(w)}{(z-w)^2}
+\frac{\p \Omg _0(w)}{(z-w)}+\cdots,  
\label{eq:omq00} \\
\Omg _2(z)\Omg _2(w)&=&\frac{c_2/2}{(z-w)^4}
+\frac{2\Omg _2(w)}{(z-w)^2}
+\frac{\p \Omg _2(w)}{(z-w)}+\cdots,
\label{eq:omq22}  \\
\Omg _0(z)\Omg _2(w)&=&\cdots, \\
T_{\psi}(z)\Omg _0(w)&=&\frac{c_0/2}{(z-w)^4}+\frac{2\Omg _0(w)}{(z-w)^2}
+\frac{\p \Omg _0(w)}{(z-w)}+\cdots,
\label{eq:Tomq0} \\
T_{\psi}(z)\Omg _2(w)&=&\frac{c_2/2}{(z-w)^4}
+\frac{2\Omg _2(w)}{(z-w)^2}
+\frac{\p \Omg _2(w)}{(z-w)}+\cdots,
\label{eq:Tomq2} \\
\Omg _0(z){\tl w} _2(w)&=&\cdots, \\
\Omg _2(z){\tl w}_2 (w)&=&\frac{1}{k+2}
\left\{ \frac{2{\tl w} _2(w)}{(z-w)^2}
+\frac{\p {\tl w} _2(w)}{(z-w)}+\cdots\right\}.
\eeqa   

\no where 

\beq
c_0=1, ~~~c_2=\frac{2(k-1)}{(k+2)}
\eeq

\no The situation here is similar to that in $W$-algebra, 
where for instance $W_3(z)$, $T(z)$ are not closed since a 
spin-$4$ field will appear in the OPE of $W_3(z)$ with itself. 
The spin-$4$ field is also decomposed into two fields which 
together with $T(z)$ and $W_3(z)$ do close to the so-called 
$W_3$ algebra~\cite{FaLy}. 

Some remarks are in order. To our knowledge, the algebraic structure 
we found is new. It is similar to a $W$-algebra structure, but now 
all the fields have conformal dimension $2$. The primary field 
${\tl w}_2 (z)$ is a ``spin-$2$ analog" of $W$-currents, and it 
transforms between fields $\psi _{a}$~(fields in Ramond sector) 
and $\psi_{\tl a}$~(fields in Neveu-Schwaz sector). In a sense, 
the primary field ${\tl w}_2(z)$, together with the other two 
spin-$2$ fields $\Omg _1(z)$ and $\Omg _2(z)$, ``non-abelianizes" 
the Virasoro algebra generated by $T(z)$. Possible applications 
of this new algebraic structure are under investigation.

In \cite{ABZam}, A.B. Zamolodchikov considered an algebra having 
$N+1$ spin-2 primary fields with OPEs given by  

\beq
T^i(z)T^j(w)=\frac{c\dl _{i,j}/2}{(z-w)^4}
+\frac{2\dl _{i,j}\dl_{j,k}T^k(w)}{(z-w)^2}
+\frac{\dl _{i,j}\dl _{j,k}\p T^k(w)}{(z-w)}
+\cdots, ~~i,~j=0,1,\cdots N, 
\eeq

\no where $T^0=T$ is the usual energy-momentum tensor. In this algebra 
there is only one central charge, and $T^i$, $i=0,~1,~\cdots,~N$,
are decoupled so that the algebra is reduced to 
$N+1$ copies of the Virasoro algebra. Our algebra is different.
Firstly, in our theory, there are 
four (independent) different central charges. So 
our theory is a conformal field theory with multi-centers. Secondly,
the four generators in our algebra are coupled to each other
and can not be reduced to copies of the Virasoro algebra.

\setcounter{section}{4}
\setcounter{equation}{0}
\section*{Acknowledgments}

We would like to express our sincere thanks to the anonymous 
referee for detailed and valuable comments and suggestions which bring 
the paper to its present form. We also thank the refree for pointing 
out an early article of A.B. Zamolodchikov~\cite{ABZam}, where
an algebra with $N+1$ decoupled spin-2 primary fields was
considered.
 
This work is financially supported by Australian Research Council. 
One of the authors (Ding) is also supported partly by the 
Natural Science Foundation of China and a grant from the AMSS.

\vskip.1in

\bebb{99}

\bibitem{ZaFa}
A.B.Zamolodchikov and V.A.Fateev, Sov. Phys, JETP {\bf 62}, 215 (1985).

\bibitem{BPZ}
A.A. Belavin, A.M. Polyakov, A.B. Zamolodchikov, {\em Nucl. Phys.} 
{\bf B241}, 333(1984).

\bbit{Hald}
F.D.M. Haldane, \prl {67}{937}{1991}.

\bibitem{FaZa}
V.A.Fateev and A.B.Zamolodchikov, Nucl.Phys.{\bf B280}, 644 (1987).

\bibitem{ZQiu}
Z.Qiu. Phys. Lett.{\bf B 198}, 497 (1987).

\bibitem{GeQi}
D.Gepner and Z.Qiu, Nucl. Phys.{\bf B285}, 423 (1987).

\bibitem{Yang}
S.K. Yang, Nucl. Phys.{\bf B285}, 639 (1987).

\bibitem{Neme}
D. Nemeschansky, Phys. Lett.{\bf B 224}, 121 (1989). 

\bibitem{DiQi}
J. Distler and Z. Qiu, Nucl. Phys.{\bf B336}, 533 (1990).

\bibitem{FrWe}
M.Freeman and P.West, Phys. Lett.{\bf B314}, 320 (1993); 
P.H. Frampton and J.T. Liu, \prl{70}{130}{1993}.

\bbit{DL}
C.Y. Dong and J. Lepowsky, {\it Generalized Vertex Algebras and 
Relative Vertex Operators}, 
Birkhauser, 1993. 

\bbit{LW}
J. Lepowsky and R. Wilson, {\em Comm. Math. Phys.} {\bf 62 }, (1978), 43-53;  
in: {\em Vertex Operators in Mathematics and Physics, 
Proc. 1983 M.S.R.I. Conference}, ed. by J. Lepowsky et. al, 
Springer-Verlag, New York, 1985, 97-142.

\bibitem{Gepn}
D. Gepner, Nucl. Phys. {\bf B290}, 10 (1987).

\bibitem{Gepn2}
D. Gepner, Phys. Lett. {\bf B199}, 380 (1987).

\bibitem{DFSW}
X.M. Ding, H. Fan, K.J. Shi, P. Wang and C.Y. Zhu, Phys. Rev. Lett. 
{\bf 70}, 2228 (1993); Nucl. Phys.{\bf B422}, 307 (1994). 

\bibitem{WaDi}
P. Wang and X.M. Ding, Phys. Lett.{\bf B335}, 56 (1994). 

\bbit{CRS}
J. M. Camino, A. V. Ramaollo and J. M. Sanchez de Santos, 
\npb{530}{715}{1998}; hep-th/9805160.

\bbit{Feld}
G. Felder, Nucl. Phys. {\bf B317}, 215 (1989).

\bbit{BerFeld}
D. Bernard and G. Felder, Commun. Math. Phys.{\bf 127}, 145 (1990).


\bbit{DGZ2}
X.M. Ding, M.D. Gould and Y.Z. Zhang,  {\it Twisted Parafermions}, 
Phys. Lett.{\bf B}, (2002), to appear; hep-th/0110165.

\bbit{Kac}
V. G. Kac, {\it Infinite Dimensional Lie Algebras}, third ed.,
Cambridge University press, Cambridge 1990.

\bbit{DGZ1}
X.M. Ding, M.D. Gould and Y.Z. Zhang, Phys. Lett.{\bf B 523}, 
367 (2001); hep-th/0109009.

\bibitem{FaLy}
V.A. Fateev and S.L. Lykyanov, Int. J. Mod. Phys. {\bf A3}, 507(1988).

\bibitem{FORT}
L. Feher, L.O'Raifeartaich, P. Ruelle, I. Tsutsui and A. Wipf, 
Phys. Rep. {\bf 222}, 1 (1992). 

\bibitem{BBSS}
F.Bais, P. Bouwknegt, M. Surridge and K. Schoutens, Nucl. Phys. {\bf B304}, 
348; 371 (1988). 

\bbit{Frnk}
E. Frenkel, {\it Free field realizations in representation theory 
and conformal field theory},  hep-th/9408109.
  
\bibitem{BoSc}
P. Bouwknegt and K. Schoutens, Phys. Rep. {\bf 223}, 183 (1993).

\bbit{Hull}
C.M.Hull, {\it Lectures on $W$-Gravity, $W$-Geometry and $W$-String}, 
hep-th-9302110.

\bibitem{West}
P. West, A review of $W$ strings, preprint 
Goteborg-ITP-93-40. 

\bbit{ABZam}
A.B. Zamolodchikov, Theor. and Math. Phys. {\bf 65}, 1205 (1985).

%

\eebb

\end{document}